\documentclass[%
 aip,
 amsmath,amssymb,
 reprint,%
]{revtex4-1}

\usepackage{graphicx}
\usepackage{dcolumn}
\usepackage{bm}

\usepackage[utf8]{inputenc}
\usepackage[T1]{fontenc}
\usepackage{mathptmx}
\usepackage{etoolbox}
\usepackage{xcolor}

\makeatletter
\def\@email#1#2{%
 \endgroup
 \patchcmd{\titleblock@produce}
  {\frontmatter@RRAPformat}
  {\frontmatter@RRAPformat{\produce@RRAP{*#1\href{mailto:#2}{#2}}}\frontmatter@RRAPformat}
  {}{}
}%
\makeatother
\begin{document}
\preprint{AIP/123-QED}

\title[]{Modeling the coincident three-ion momentum imaging of diiodomethane photodissociation on reduced-dimensional potential energy surfaces}
\author{Yijue Ding}
 \affiliation{Department of Chemistry, Southern University of Science and Technology, Shenzhen, Guangdong, China}
 \email{yijueding@gmail.com}



\begin{abstract}
We present an efficient theoretical model to simulate observables in the time-resolved coincident three-ion Coulomb explosion experiment of diiodomethane. The model employs two degrees of freedom to describe the C–I bond breaking and the $\text{CH}_2\text{I}$ rotation during photodissociation, and three degrees of freedom to describe the coincident $\text{CH}_2^{+} + \text{I}^{2+} + \text{I}^{2+}$ fragmentation during the subsequent Coulomb explosion. By solving the equations of motion, the photodissociation pathways are obtained on two-dimensional potential energy surfaces of the valence excited states of the neutral molecule, and the asymptotic momenta of the three ionic fragments are determined on the three-dimensional ground-state potential energy surface of the fivefold-charged cation. The photodissociation pathways are consistent with previous \textit{ab initio} molecular dynamics simulations and indicate a $\text{CH}_2\text{I}$ rotational period of approximately 340 fs. The theoretical time-resolved kinetic energy release and the correlation between the kinetic energy release and the angle between the two $\text{I}^{2+}$ momenta show good agreement with experimental signals in part, reflecting and confirming the static $\text{CH}_2\text{I}_2$ state and the $\text{CH}_2\text{I} + \text{I}$ dissociation channels.
\end{abstract}

\maketitle

\section{Introduction}

Following molecular structural changes during chemical reactions is vital for understanding reaction mechanisms, which typically requires ultrafast imaging technology. Ultrafast molecular imaging methods are usually categorized into two groups: direct and indirect imaging methods.
Indirect imaging methods, such as transient absorption spectroscopy (TAS)\cite{ftas,atas} and time-resolved photoelectron spectroscopy (TRPES)\cite{trpes}, are based on electron spectroscopy and rely on sophisticated electronic-structure theory to retrieve molecular structures. Direct imaging methods, such as ultrafast electron diffraction (UED)\cite{ued} and Coulomb explosion imaging (CEI)\cite{reviewcei}, measure nuclear dynamics and retrieve nuclear coordinates in a more straightforward manner. CEI has been widely used due to its good temporal resolution and its equal sensitivity to both light and heavy atoms.
In modern laboratories, CEI is typically realized through laser-induced photoionization, in which an intense laser pulse strips multiple electrons from a neutral molecule, resulting in strong Coulomb repulsion among the ionic fragments. The final momentum of each ionic fragment is then measured using velocity map imaging (VMI)\cite{burt2017,burt2018,dingdajun2017,amini2018} or cold target recoil-ion momentum spectrometer (COLTRIMS)\cite{vager1989,boll2022,endo2020} setups. Due to the low count rate in multi-fragment coincidence detection, time-resolved CEI is usually limited to two-body or three-body fragmentation channels\cite{cornaggia2009,yatsuhashi2018,li2022b}.

As a direct imaging method, CEI retrieves molecular structures based on the measured ionic momenta. Thus, it requires a reliable theoretical framework that maps these momenta to the geometric structure of a molecule. However, developing a rigorous theory of the Coulomb explosion process is challenging. First, the strong-field ionization or Auger–Meitner decay processes that drive the molecule to highly ionized states are difficult to model. Second, the interactions between ionic fragments are not purely Coulombic, and the exact ionic states to which the molecule evolves remain unknown\cite{wright1999,schouder2020}.
Moreover, models that assist time-resolved CEI in probing photochemical reactions typically involve intricate multi-dimensional molecular dynamics (MD) simulations\cite{ding2024,ding2025b,endo2020,zhou2020}. Whether performed on-the-fly or on pre-built potential energy surfaces, MD simulations require enormous computational resources.

In this work, we develop a theoretical model to simulate time-resolved coincident three-fragment Coulomb explosion, extending our previous work on modeling two-fragment CEI\cite{ding2025a}. Our goal is to reduce the dimensionality of the problem while still capturing the most important physics during photodissociation and Coulomb explosion. Unlike two-fragment CEI, which provides only one-dimensional information (usually the dissociative bond length) about the transient molecular structure, coincident three-fragment CEI provides multidimensional information that may indicate multiple reaction channels. As an example, we apply our theory to the time-resolved three-fragment Coulomb explosion of diiodomethane ($\text{CH}_2\text{I}_2$), in analogy to a recent pump-probe experiment\cite{anbu2025}. Specifically, the experiment uses a 290 nm UV pump pulse to trigger photodissociation and an 800 nm strong IR probe pulse to ionize the molecule to the fivefold charged channel, leading to the $\text{CH}_2^{+} + \text{I}^{2+} + \text{I}^{2+}$ three-body fragmentation. With our theory, we aim to provide a direct comparison with the experimental observables and confirm several reaction channels suggested by the experiment.

This work is organized as follows. In Section II, we elaborate on the construction of reduced-dimensional potential energy surfaces for the neutral molecule and the fivefold charged cation. In Sections III and IV, we present our theoretical framework and apply it to simulate the photodissociation and three-fragment Coulomb explosion processes, as well as to compare our results with the CEI observations. Finally in Section V, we draw a brief conclusion summarizing our study.

\section{Potential energy surfaces in reduced dimension}

\begin{figure}
    \centering
    \includegraphics[width=0.8\linewidth]{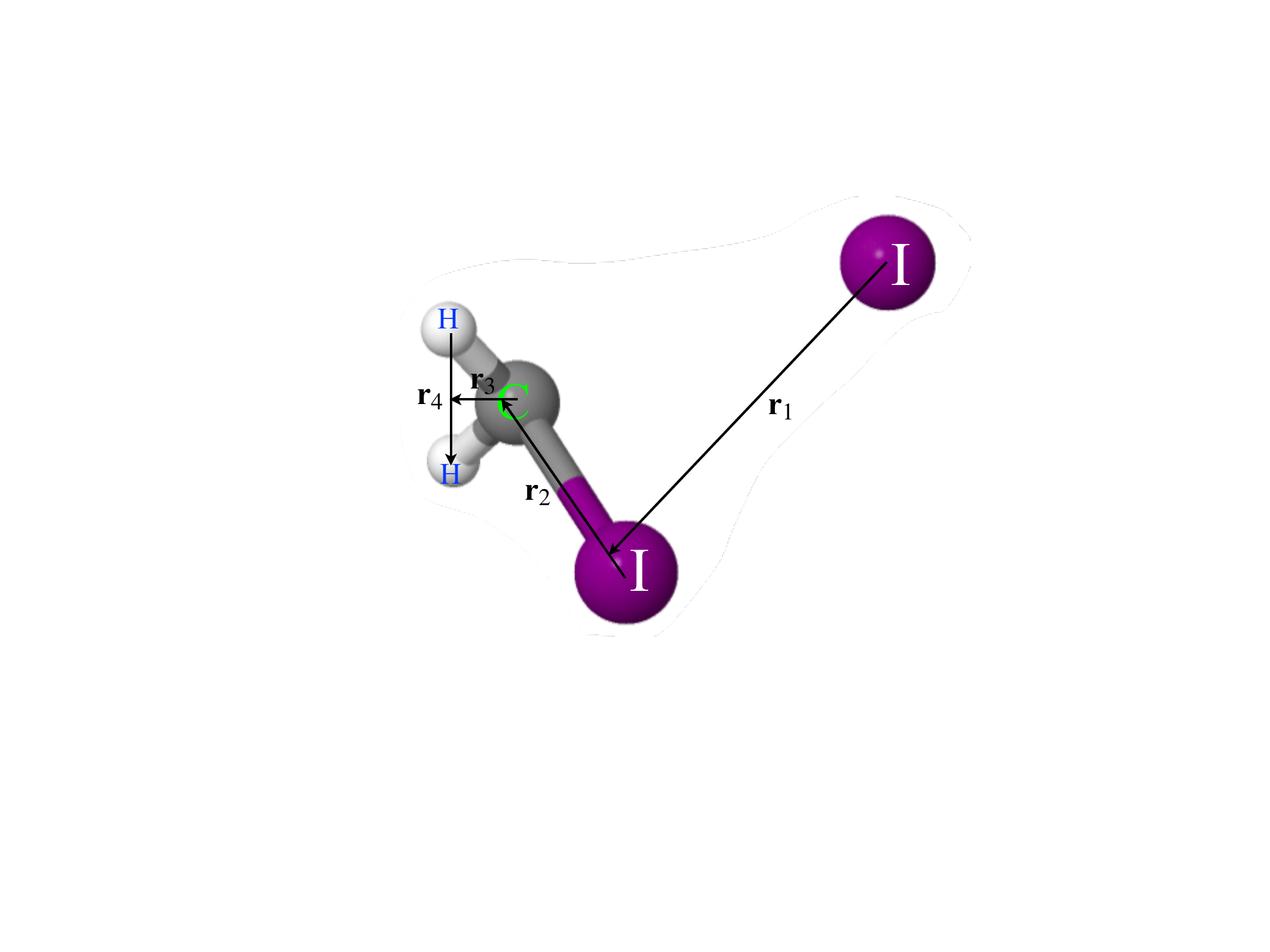}
    \caption{Molecular diagram of $\text{CH}_2\text{I}_2$ in $C_s$ symmetry. $\textbf{r}_i(i=1\dots4)$ denote the Jacobi coordinates employed to model the photodissociation and Coulomb explosion processes. }
    \label{fig1}
\end{figure}

We propose using a minimal number of degrees of freedom (DOF) to model the photodissociation and the subsequent Coulomb explosion. During photodissociation, one C–I bond breaks while inducing rotation of the $\text{CH}_2\text{I}$ fragment. Therefore, at least two internal DOF must be considered to characterize both the C–I translational motion and the rotation of the $\text{CH}_2\text{I}$ fragment.
For the Coulomb explosion leading to $\text{CH}_2^{+}$, $\text{I}^{2+}$, and $\text{I}^{2+}$ ionic fragments, at least three internal DOF must be included in the reaction coordinates to describe such three-body fragmentation. To this end, we employ Jacobi coordinates to model these two processes, as illustrated in the molecular diagram in Fig.~\ref{fig1}. We select $r_1$ (the magnitude of $\mathbf{r}_1$ vector) to characterize the C–I bond breaking and $\alpha$ (the angle between the $\mathbf{r}_1$ and $\mathbf{r}_2$ vectors) to describe the rotation of the $\text{CH}_2\text{I}$ component during photodissociation, while constraining all other DOF to their equilibrium geometry values. In addition to $r_1$ and $\alpha$, we include $r_2$ (the magnitude of $\mathbf{r}_2$ vector) in the reaction coordinates to characterize the $\text{CH}_2^{+}$–$\text{I}^{2+}$ separation during the three-fragment Coulomb explosion. In the following subsections, we elaborate on the construction of the two-dimensional potential energy surface (PES) of neutral diiodomethane and the three-dimensional PES of its fivefold charged cation.

\subsection{Neutral molecule ($\text{CH}_2\text{I}_2$)}

\begin{table}
\caption{Optimized geometric parameters (length in angstrom and angle in degree) of $\text{CH}_2\text{I}_2$ at equilibrium at the MRCI(12,8)/cc-pVTZ level of theory.}
\begin{ruledtabular}
\begin{tabular}{cc}
 Parameter & Optimized value  \\
\hline
C-I length & 2.15			 		\\
I-I length  & 3.64	 		 \\
C-H length  & 1.08	 		 \\
I-C-H angle & 107.6	  		\\
H-C-H angle	& 112.2	 		\\
I-C-I angle	& 115.2	 		\\
\end{tabular}
\end{ruledtabular}
\label{tab1}
\end{table}

The $\text{CH}_2\text{I}_2$ electronic structure is calculated at the multi-reference configuration interaction (MRCI)\cite{mrci1,mrci2} level of theory using an active space of 12 electrons in 8 orbitals. A state-averaged multi-configuration self-consistent field (SA-MCSCF)\cite{mcscf1,mcscf2} calculation is first performed to obtain the reference wave functions. The MRCI calculation is then carried out based on the SA-MCSCF reference to further optimize the electronic states by including single and double excitations to external orbitals. Spin–orbit (SO) coupling effects are treated using the state-interaction method. In this approach, the SO-coupled Hamiltonian matrix is constructed in the basis of MCSCF wave functions, with the diagonal elements replaced by the corresponding MRCI energies. The SO-coupled eigenstates are then obtained by diagonalizing this Hamiltonian. For all electronic structure calculations, we employ the cc-pVTZ basis set for hydrogen and carbon atoms and the cc-pVTZ-PP basis set for iodine\cite{dunningbasis}. In the cc-pVTZ-PP basis set, the innermost 28 electrons of the iodine atom are replaced by a relativistic pseudopotential\cite{pseudopotential,pseudopotential2}. All electronic structure calculations are performed using the MOLPRO quantum chemistry package\cite{molpro,molpro2}.

\begin{figure}
    \centering
    \includegraphics[width=\linewidth]{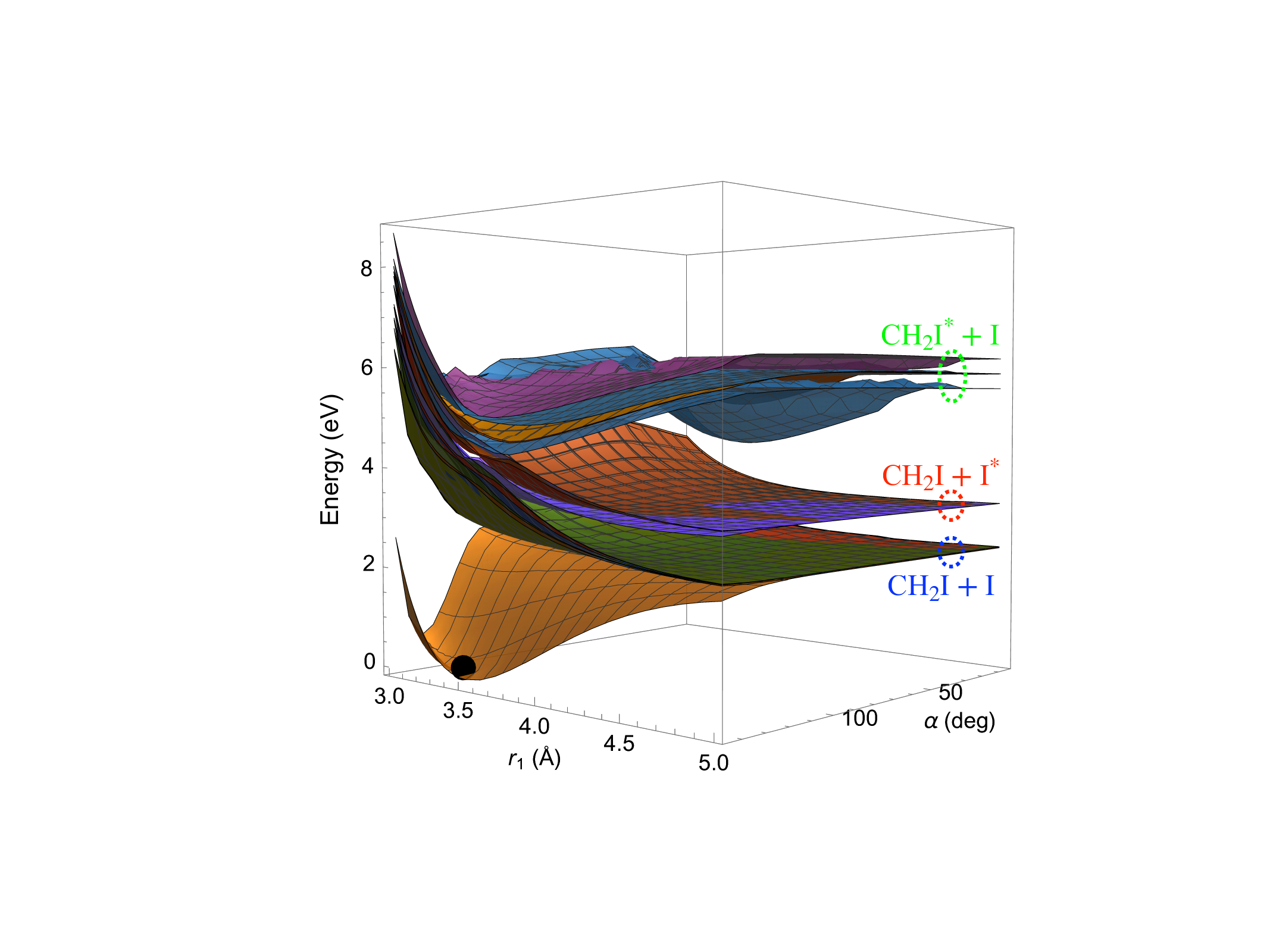}
    \caption{Potential energies of the $\text{CH}_2\text{I}_2$ molecule as functions of $r_1$ and $\alpha$ in Jacobi coordinates for the lowest 17 electronic states. These states can be grouped according to their corresponding dissociation thresholds, as indicated in the figure. The black dot marks the equilibrium geometry, which also corresponds to the Franck-Condon point upon photoexcitation.}
    \label{fig2}
\end{figure}

We calculate five singlet states and four triplet states of $\text{CH}_2\text{I}_2$, resulting in a total of 17 SO-coupled states. These states include the ground state and all excited states that are accessible via single-photon absorption near 290 nm. To construct the PES, geometry optimizations are first performed at the ground-state MRCI level to obtain the equilibrium geometric parameters of the molecule, as listed in Table~\ref{tab1}. Electronic structure calculations are then carried out on a structured geometric grid of $\{r_1,\alpha\}$, with the remaining DOF constrained to their equilibrium geometry values. Finally, the \textit{ab initio} energy points are interpolated using cubic spline functions to obtain an analytic representation of the potential energy, $V = V(r_1,\alpha)$. 






Figure \ref{fig2} shows the interpolated PES for the 17 calculated electronic states. A significant number of excited states are nearly degenerate, possibly due to the geometric constraints that enforce $C_s$ symmetry. It should be noted that with the SO coupling, the spatial symmetry ($A'$ and $A''$) is no longer a good quantum number of the electron wavefunction. Instead, we can use spinor symmetry to identify electronic states, labeled as $\Gamma_1$ and $\Gamma_2$.
These states dissociate into (1) $\text{CH}_2\text{I}+\text{I}$, (2) $\text{CH}_2\text{I}+\text{I}^*$, and (3) $\text{CH}_2\text{I}^*+\text{I}$ asymptotes at large $r_1$. Here, $\text{I}(^2\text{P}_{3/2})$ and $\text{I}^*(^2\text{P}_{1/2})$ correspond to the two atomic states of iodine with a energy splitting of 0.9 eV, and $\text{CH}_2\text{I}^*$ denotes the excited state of $\text{CH}_2\text{I}$ radical. The present calculation does not reveal the SO splitting in the $\text{CH}_2\text{I}$ radical. The ground-state equilibrium geometry is located at $r_1=3.46~\text{ \AA}$ and $\alpha=143.3^\circ$. This geometry is also the Frack-Condon (FC) point when the molecule vertically transitions to excited states upon photoexcitation. 
Several avoided crossings are present near the FC region. Previous studies have investigated nonadiabatic transitions between the group (2) and group (3) states mediated by these crossings, which give rise to so-called indirect dissociation pathways\cite{liu2020}. In the present work, however, nonadiabatic processes are not considered.



\subsection{Fivefold-charged cation ($\text{CH}_2\text{I}_2^{5+}$)}

\begin{figure*}
\includegraphics[width=0.8\textwidth]{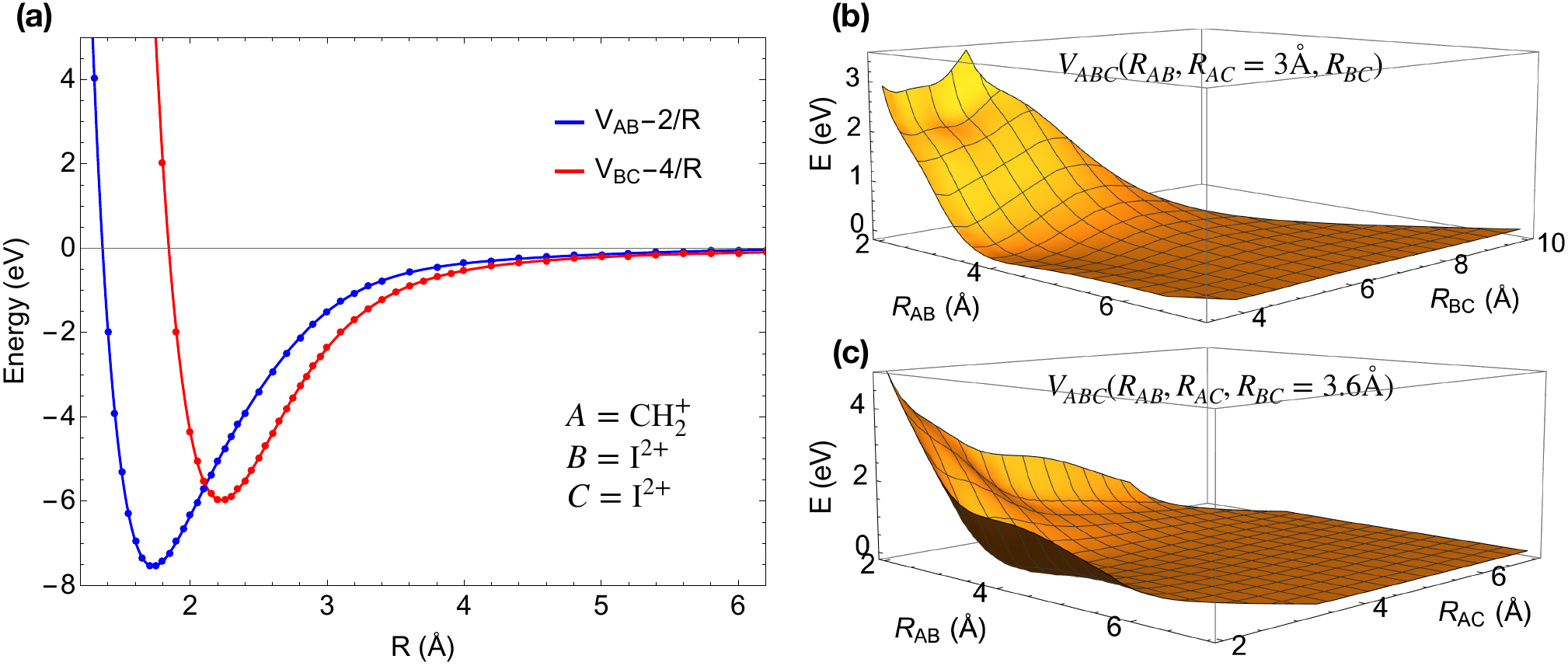}
\caption{Two-body and three-body interaction potentials for the ground-state potential energy surface of the $\text{CH}_2\text{I}_2^{5+}$ cation. (a) The two-body interaction potentials $V_{AB}$ (blue) and $V_{BC}$ (red), with the pure Coulomb contribution subtracted, shown as a function of the pair distance $R$. The dots represent the original \textit{ab initio} energies, and the curves correspond to the fitted potential function. (b) The three-body interaction potential $V_{ABC}$ as a function of $R_{AB}$ and $R_{BC}$, with $R_{AC}$ fixed at 3 \AA. (c) $V_{ABC}$ as a function of $R_{AB}$ and $R_{AC}$, with $R_{BC}$ fixed at 3.6 \AA. Note that only physically allowed $V_{ABC}$ values--i.e. those for which $R_{AB}$, $R_{AC}$, and $R_{BC}$ satisfy the triangle inequality--are shown in panels (b) and (c).}
\label{fig3}
\end{figure*} 


The ionic PES is constructed using the many-body expansion approach\cite{aguado1992,aguado1998}. In this method, the potential energy function is expressed as a sum of two-body and three-body interaction terms, given by
\begin{equation}
V_{\text{ion}} = V_{AB} + V_{AC} + V_{BC} + V_{ABC},
\end{equation}
where $A$, $B$, and $C$ represent the three ionic fragments (in this study, $\text{CH}_2^{+}$, $\text{I}^{2+}$, and $\text{I}^{2+}$, respectively).



The two-body interaction functions $V_{AB}$ and $V_{AC}$, which represent the interactions between the $\text{CH}_2^{+}$ and $\text{I}^{2+}$ fragments, are approximately equivalent. We perform ground-state MRCI calculations for the $\text{CH}_2\text{I}^{3+}$ cation using an active space of 4 electrons in 5 orbitals on a grid of the distance $R$ between the $\text{I}^{2+}$ fragment and the center of mass (CM) of the $\text{CH}_2^{+}$ fragment, with the remaining DOF fixed at their values of the $\text{CH}_2\text{I}_2$ equilibrium geometry. Subsequently, the \textit{ab initio} energies are fitted to an analytic expression given by
\begin{equation}
V_{AB}(R_{AB})=\frac{q_A q_B}{R_{AB}}-\frac{C_4}{R_{AB}^4} e^{-\beta_1 R_{AB}}+\sum_{i=1}^{n} c_i \left(R_{AB} e^{-\beta_2 R_{AB}}\right)^i,
\label{eq2}
\end{equation}
where the first term represents the pure Coulomb interaction, the second term is a damped long-range form that accounts for the charge-dipole interaction, and the remaining terms describe the short- to intermediate-range covalent interaction and ensure that $V_{AB} \rightarrow 0$ as $R_{AB} \rightarrow \infty$.

Likewise, the two-body interaction function $V_{BC}$, which represents the interaction between the two $\text{I}^{2+}$ fragments, is constructed using the same approach. We first perform ground-state MRCI calculations for the $\text{I}_2^{4+}$ cation using an active space of 6 electrons in 6 orbitals on a grid of the distance $R_{BC}$ between the two $\text{I}^{2+}$ fragments. $V_{BC}$ is then fitted to the \textit{ab initio} energy points using the same functional form as $V_{AB}$ in Eq. \eqref{eq2}.

The fitted two-body interaction functions $V_{AB}$ and $V_{BC}$ are shown in Fig. \ref{fig3}(a) and exhibit excellent agreement with the \textit{ab initio} energy points. The root mean square error (RMSE) of both fittings is extremely small ($<1$ meV). The attractive covalent interactions between the ionic fragments are strong at short distances, but as the inter-fragment distance increases, the interaction becomes dominated by Coulomb repulsion and is nearly purely Coulombic for $R > 6$ \AA.

To construct the three-body interaction function $V^{}_{ABC}$, we first perform the ground-state MRCI calculation for $\text{CH}_2\text{I}_2^{5+}$ cation using an active space of 7 electrons in 8 orbitals on a structured grid of $\{R_{AB},R_{AC},R_{BC}\}$. 
The energy data for the three-body interaction term are obtained by subtracting the fitted two-body interaction energies from the \textit{ab initio} energies of the $\text{CH}_2\text{I}_2^{5+}$ cation. These data are then fitted to a polynomial function, written as
\begin{equation}
    V_{ABC}(R_{AB},R_{AC},R_{BC})=\sum_{i,j,k=1}^MC_{ijk}\rho_{AB}^i\rho_{AC}^j\rho_{BC}^k,
    \label{eq3}
\end{equation}
where $\rho_{AB}$,$\rho_{AC}$, and $\rho_{BC}$ are defined as 
\begin{equation}
    \begin{split}
        \rho_{AB}=R_{AB}e^{-\gamma_{AB}R_{AB}},\\
        \rho_{AC}=R_{AC}e^{-\gamma_{AC}R_{AC}},\\
        \rho_{BC}=R_{BC}e^{-\gamma_{BC}R_{BC}}.
    \end{split}
\end{equation}
In Eq. \eqref{eq3}, the indices are restricted to satisfy $i+j+k\le M$. Specifically, we choose $M=8$, which results in 56 linear parameters $C_{ijk}$ and 3 non-linear parameters $\gamma_{AB}$,$\gamma_{AC}$, and $\gamma_{BC}$.

The three-body interaction function $V_{ABC}$ is obtained with reasonable accuracy, which a RMSE of approximately 70 meV. Figure \ref{fig3}(b) and \ref{fig3}(c) show two cuts of $V_{ABC}$ to demonstrate the fit quality and to illustrate the property of the three-body interaction. The three-body interaction is repulsive when all three inter-fragment distances are small, compensating for the two-body covalent attraction. $V_{ABC}$ decays rapidly when any of the inter-fragment distances increases, indicating that the three-body term is a short-range interaction. 



\section{Photodissociation pathways}

In our theoretical model, only two internal DOF, $r_1$ and $\alpha$, are included in the photodissociation reaction coordinates. These two DOF capture the most significant structural changes during photodissociation: the C–I bond breaking (characterized by $r_1$) and the $\text{CH}_2\text{I}$ rotation (characterized by $\alpha$). Therefore, the $\text{CH}_2\text{I}$ fragment is treated as a rigid body. Moreover, the geometric constraints imposed on the molecule ensure that the rotation occurs only in the $\text{CI}_2$ plane. Therefore, it is convenient to use polar coordinates in the $\text{CI}_2$ plane to describe the equations of motion (EOM).
The selected DOF to model the photodissociation are ${r_1, \theta_1, \theta}$, where $\theta_1$ and $\theta$ are the polar angles of the $\mathbf{r}_1$ and $\mathbf{r}_2$ vectors, respectively. The internal angle $\alpha$ can be expressed as $\alpha = \theta_1 - \theta$.
The total kinetic energy of the system is given by
\begin{equation}
   T=\frac{1}{2}\sum_{i=1}^4\mu_i\dot{\textbf{r}}_i^2=\frac{1}{2}\mu_1(\dot{r}_1^2+r_1^2\dot{\theta}_1^2)+\frac{1}{2}(\mu_2 r_2^2+\mu_3 r_3^2)\dot{\theta}^2 
\end{equation}
where $\mu_i(i=1\dots 4)$ is the reduced mass and $\dot{q}=dq/dt(q=\{r_1, \theta_1,\theta\})$ denotes the generalized velocity. Using the Euler-Lagrange formalism, the EOM is written as 
\begin{equation}
\begin{split}
    \mu_1\ddot{r}_1-\mu_1r_1\dot{\theta}_1^2=-\frac{\partial V}{\partial r_1}, \\
    \mu_1r_1^2\ddot{\theta}_1+2\mu_1r_1\dot{r}_1\dot{\theta}_1=-\frac{\partial V}{\partial \theta_1}, \\
   (\mu_2r_2^2+\mu_3r_3^2)\ddot{\theta}=\frac{\partial V}{\partial \theta_1}.
\end{split}
\label{eq6}
\end{equation} 

Because the molecule is likely to transition to multiple excited states upon UV excitation, we select the $2\Gamma_1$ (2nd energy state in $\Gamma_1$ spinor symmetry) and $6\Gamma_2$ (6th energy state in $\Gamma_2$ spinor symmetry) adiabatic states, corresponding to the $\text{CH}_2\text{I} + \text{I}$ and $\text{CH}_2\text{I} + \text{I}^*$ reaction channels, as representative examples to study the photodissociation. In our theoretical model, the molecule starts at the FC point with zero initial velocity on the $2\Gamma_1$ and $6\Gamma_2$ states and moves adiabatically on each PES toward dissociation. The reaction paths are obtained by integrating the EOM in Eq. \eqref{eq6}. Although avoided crossings are present near the FC point, nonadiabatic transitions are not considered in the present study.

\begin{figure}
    \centering
    \includegraphics[width=\linewidth]{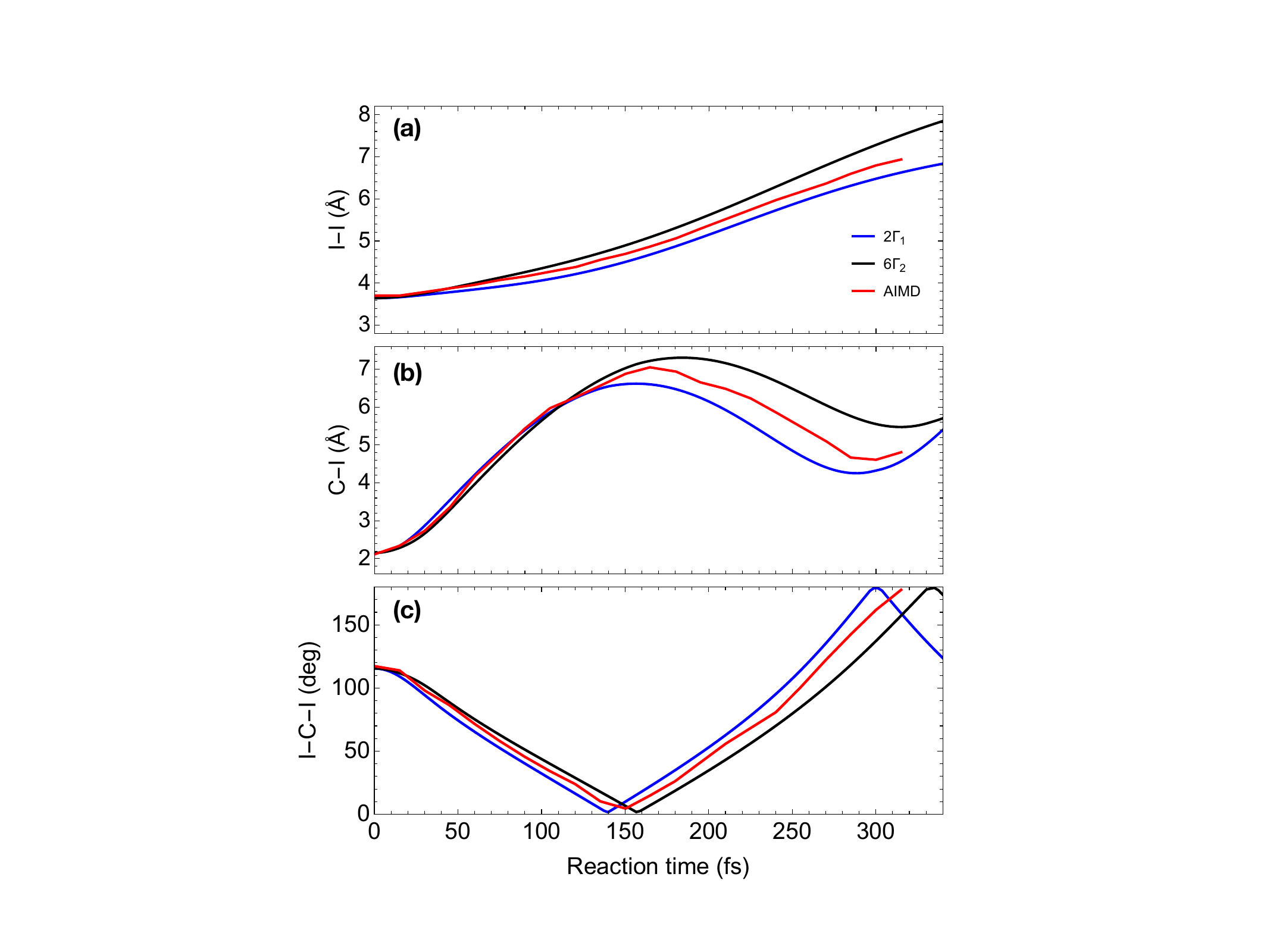}
    \caption{The I--I (a) and C--I (b) pair distances and the I--C--I bond angle (c) as functions of reaction time along the photodissociation pathways for the $2\Gamma_1$ (blue) and $6\Gamma_2$ (black) states, which lead to the $\text{CH}_2\text{I} + \text{I}$ and $\text{CH}_2\text{I} + \text{I}^*$ products, respectively. Results from a representative AIMD trajectory (red, reproduced from Ref. \onlinecite{liu2020}) are also shown for comparison. The AIMD results are reproduced with permissions from Phys. Rev. X \textbf{10}, 021016 (2020). Copyright 2020 American Physical Society. }
    \label{fig4}
\end{figure}

The time-resolved I–I and C–I pair distances, as well as the I–C–I bond angle along these reaction paths, are shown in Fig.~\ref{fig4} and are compared with a representative \textit{ab initio} molecular dynamics (AIMD) trajectory\cite{liu2020}. The AIMD calculation evaluates the electronic structure on-the-fly using the CASPT2(12,8)/ANO-RCC-VDZP method and includes all DOF of the molecule. The reaction paths predicted by our theoretical model show overall agreement with the AIMD trajectory, confirming the validity of dimensionality reduction during photodissociation.
Our predicted reaction paths and the AIMD trajectory exhibit the same qualitative behavior for these quantities: the I–I distance increases monotonically with reaction time; the C–I distance shows an overall increasing trend accompanied by oscillations arising from the rotation of the $\text{CH}_2\text{I}$ radical; and the oscillations in the I–C–I angle further indicate that the $\text{CH}_2\text{I}$ rotation is excited during photodissociation. The $\text{CH}_2\text{I}$ rotation has a period of approximately 340 fs, in agreement with recent UED and CEI measurements\cite{liu2020,anbu2025}.

\section{Three-fragment Coulomb explosion}
To simulate experimental observables, we focus on the Coulomb explosion of the fivefold-charged cation $\text{CH}_2\text{I}_2^{5+}$, which breaks up into three ionic fragments: $\text{CH}_2^{+}$, $\text{I}^{2+}$, and $\text{I}^{2+}$. In the present study, we do not model the strong-field–induced multifold ionization in detail; instead, we assume that the strong IR pulse instantaneously removes five valence electrons from the neutral molecule, leaving the nuclear motion unchanged. Consequently, the reaction time during photodissociation is treated as the pump–probe delay.
Furthermore, the exact electronic state of the $\text{CH}_2\text{I}_2^{5+}$ cation remains unknown during Coulomb explosion. Therefore, in this work we only consider Coulomb explosions on the ground state of $\text{CH}_2\text{I}_2^{5+}$ cation as an example. 

In the COLTRIMS experiment, the coincidence of the three ionic fragment is ensured by momentum conservation:
\begin{equation}
    \textbf{p}_A+\textbf{p}_B+\textbf{p}_C=0,
    \label{eq7}
\end{equation}
where $\textbf{p}_A$, $\textbf{p}_B$, and $\textbf{p}_C$ are the asymptotic momenta of each ionic fragment. In our simulation, these momenta are calculated with respect to the CM of the cation, which always ensures Eq. \eqref{eq7}. 

As discussed in Section II, since the center of charge is approximated to coincide with the CM, the Coulomb explosion does not induce further rotations of the ionic fragments. In Section II, the ionic potential energy is expressed as a function of the inter-fragment distances $\{R_{AB}, R_{AC}, R_{BC}\}$, but we still use the Jacobi coordinates $\{r_1, r_2, \alpha\}$ to model the three-fragment Coulomb explosion. The transformation between these coordinates is given by
\begin{equation}
\begin{split}
     R_{AB}=r_2, \\
     R_{AC}=\sqrt{r_1^2+(m_B/m_{AB})^2r_2^2+2r_1(m_{B}/m_{AB})r_2\cos\alpha}, \\
     R_{BC}=\sqrt{r_1^2+(m_A/m_{AB})^2r_2^2-2r_1(m_{A}/m_{AB})r_2\cos\alpha}.
\end{split}
\label{eq8}
\end{equation}

Similar to the procedure used in photodissociation, we select four DOF $\{r_1, r_2, \theta_1, \theta\}$ to model the three-fragment Coulomb explosion. In this case, the total kinetic energy is given by 
\begin{equation}
    T=\frac{1}{2}\mu_1(\dot{r}_1^2+r_1^2\dot{\theta}_1^2)+\frac{1}{2}\mu_2( \dot{r}_2^2+r_2^2\dot{\theta}^2)+\frac{1}{2}\mu_3 r_3^2\dot{\theta}^2.
    \label{eq9}
\end{equation}
The EOM in Euler-Lagrange formalism is given by
\begin{equation}
\begin{split}
    \mu_1\ddot{r}_1-\mu_1r_1\dot{\theta}_1^2=-\frac{\partial V_\text{ion}}{\partial r_1}, \\
    \mu_1r_1^2\ddot{\theta}_1+2\mu_1r_1\dot{r}_1\dot{\theta}_1=-\frac{\partial V_\text{ion}}{\partial \theta_1}, \\
   (\mu_2r_2^2+\mu_3r_3^2)\ddot{\theta}+2\mu_2r_2\dot{r}_2\dot{\theta}=\frac{\partial V_\text{ion}}{\partial \theta_1}\\
   \mu_2\ddot{r}_2-\mu_2r_2\dot{\theta}^2=-\frac{\partial V_\text{ion}}{\partial r_2}.
\end{split}
\label{eq10}
\end{equation}
Therefore, the asymptotic momentum of each ionic fragment is obtained by integrating Eq. \eqref{eq9} until $V_\text{ion}<0.1$ eV. 

\begin{figure}
    \centering
    \includegraphics[width=\linewidth]{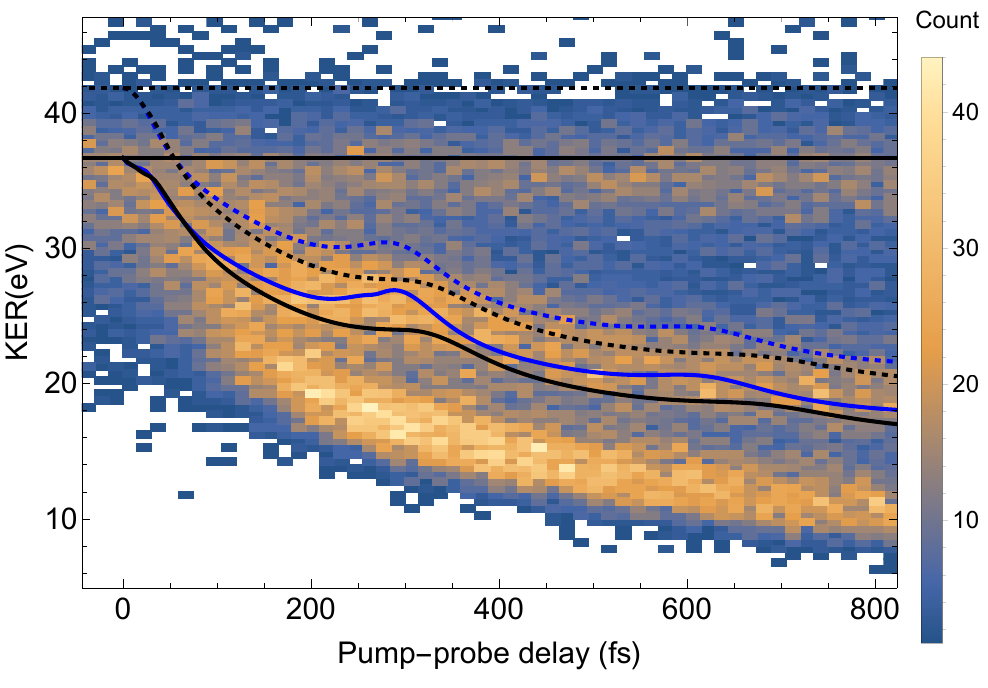}
    \caption{Experimental KER signals at different pump-probe delays (reproduced from Ref. \onlinecite{anbu2025}) compared with theoretical predictions. The experimental KER is the sum of the kinetic energies of the three ionic fragments, $\text{CH}_2^{+}$, $\text{I}^{2+}$, and $\text{I}^{2+}$, measured in coincidence. The curves represent the theoretical results. Photodissociation is modeled using the $2\Gamma_1$ (blue) and $6\Gamma_2$ (black) states that lead to $\text{CH}_2\text{I}+\text{I}$ and $\text{CH}_2\text{I}+\text{I}^*$ dissociation channels, respectively. The subsequent three-fragment Coulomb explosion is modeled using the ground-state PES of the $\text{CH}_2\text{I}_2^{5+}$ cation (solid curves) and, for comparison, a pure Coulomb potential (dashed curves). The experimental data are reproduced with permissions from J. Chem. Phys. \textbf{163}, 164308 (2025). Copyright 2025 American Institute of Physics.}
    \label{fig5}
\end{figure}

The final kinetic energy release (KER) is a key observable in the Coulomb explosion experiment, it is written as the sum of each ionic fragment:
\begin{equation}
    \mathrm{KER}=\frac{\textbf{p}_A^2}{2m_A}+\frac{\textbf{p}_B^2}{2m_B}+\frac{\textbf{p}_C^2}{2m_C}
\end{equation}
Figure \ref{fig5} shows the experimental time-resolved KER compared with our theoretical predictions. The experimental KER signals can be grouped into three bands: a delay-independent band centered around 36 eV, an upper delay-dependent band decreasing to about 19 eV at 800 fs, and a lower delay-dependent band decreasing to about 10 eV at 800 fs.
The theoretical KER is simulated for the $2\Gamma_1$ and $6\Gamma_2$ dissociation states using both the full ionic potential and a pure Coulomb potential. Overall, the theoretical delay-dependent KER aligns well with the upper delay-dependent band of the experimental signals, confirming that these signals correspond to the $\text{CH}_2\text{I} + \text{I}$ dissociation channel.
The theoretical KER increases slightly near 300 fs and 600 fs due to the rotation of the $\text{CH}_2\text{I}$ fragment, which leads to a temporary decrease in the C–I distance during photodissociation. The KER simulated using the full ionic potential is approximately 4 eV lower than that obtained with a pure Coulomb potential and shows better agreement with the experimental signals, indicating the importance of non-Coulombic effects during the Coulomb explosion.
Moreover, the experimental delay-independent signals are attributed to static $\text{CH}_2\text{I}_2$ near its equilibrium geometry. Our theory confirms this by simulating the Coulomb explosion of $\text{CH}_2\text{I}_2$ at its equilibrium geometry (i.e., at zero pump-probe delay) using the full ionic potential, whereas the result obtained with a pure Coulomb potential is about 5 eV higher than the center of the experimental delay-independent band.
The lower delay-dependent band of the experimental signals could possibly be attributed to the $\text{CH}_2 + \text{I} + \text{I}$ dissociation channel, although this has not been confirmed yet.



Besides KER, the directions of the momenta of the ionic fragments also carry important structural information about the molecule. The correlation between KER and the ionic momentum vector angle is commonly used to identify reaction channels\cite{endo2020,anbu2025}. In the coincident three-ion Coulomb explosion experiment conducted on $\text{CH}_2\text{I}_2$, analysis of the KER-angle correlation has led researchers to propose that multiple reaction pathways exist upon UV excitation, including $\text{CH}_2\text{I}+\text{I}$ dissociation, $\text{CH}_2+\text{I}+\text{I}$ dissociation, $\text{CH}_2+\text{I}_2$ elimination, and $\text{CH}_2\text{I--I}$ isomerization\cite{anbu2025}. However, most of these proposed reaction channels still await confirmation from theory.

\begin{figure}
    \centering
    \includegraphics[width=\linewidth]{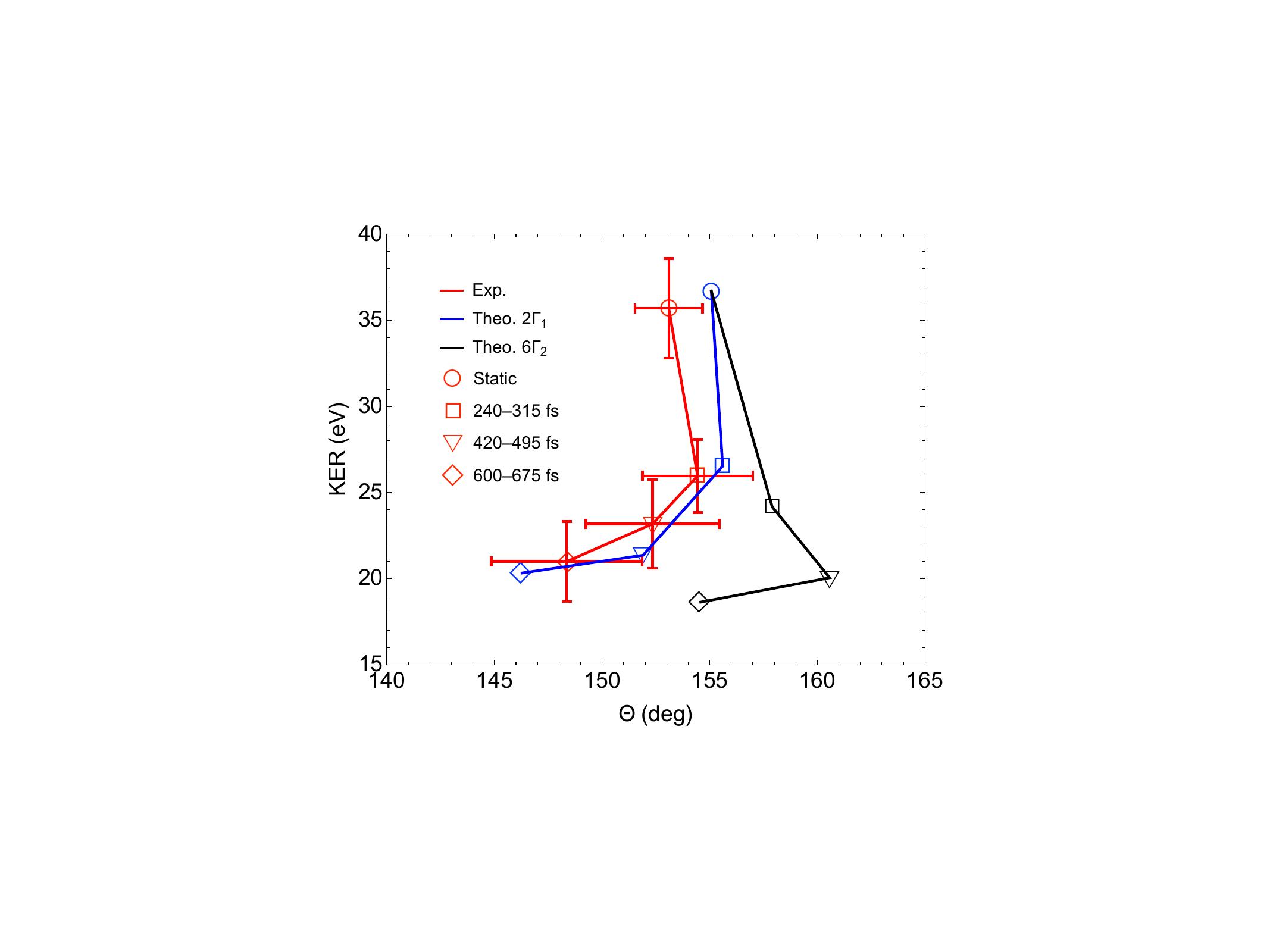}
    \caption{The correlation between KER and the angle $\Theta$ between the asymptotic momenta of the two $\text{I}^{2+}$ cations, obtained from the coincident three-ion ($\text{CH}_2^{+}$, $\text{I}^{2+}$, and $\text{I}^{2+}$) Coulomb explosion experiment (reproduced from Ref. \onlinecite{anbu2025}) and from theoretical predictions. Results are shown for the static $\text{CH}_2\text{I}_2$ molecule (circles) and for the transient $\text{CH}_2\text{I}+\text{I}$ dissociation state at different pump-probe delay intervals (squares: 240--315 fs; triangles: 420--495 fs; diamonds: 600--675 fs). The experimental results (red) display the mean values (markers) and the standard deviations (error bars) of KER and $\Theta$ for those coincidence signals corresponding to the static $\text{CH}_2\text{I}_2$ and the $\text{CH}_2\text{I}+\text{I}$ dissociation channel.
    The theoretical results show the average values in each pump-probe delay interval. The theoretical modeling is performed using the $2\Gamma_1$ (blue) and $6\Gamma_2$ (black) states for photodissociation and the ground-state PES of the $\text{CH}_2\text{I}_2^{5+}$ cation for the subsequent Coulomb explosion.  
    The experimental data are reproduced with permissions from J. Chem. Phys. \textbf{163}, 164308 (2025). Copyright 2025 American Institute of Physics.}
    \label{fig6}
\end{figure}

Although our theory has successfully identified the static $\text{CH}_2\text{I}_2$ state and the $\text{CH}_2\text{I}+\text{I}$ dissociation channel based on the KER alone, we further support these assignments by analyzing the KER-angle correlation and comparing it with the experimental results. In the coincident $\text{CH}_2^{+}+\text{I}^{2+}+\text{I}^{2+}$ Coulomb explosion, we analyze the angle $\Theta$ between the asymptotic momenta of the two $\text{I}^{2+}$ cations, defined as
\begin{equation}
    \cos\Theta=\frac{\textbf{p}_B\cdot\textbf{p}_C}{|\textbf{p}_B||\textbf{p}_C|}.
\end{equation}
Figure \ref{fig6} shows the experimental KER-$\Theta$ correlation compared with the theoretical predictions at different pump-probe delay intervals. Since we focus only on the static $\text{CH}_2\text{I}_2$ state and the $\text{CH}_2\text{I}+\text{I}$ dissociation channel, experimental signals corresponding to other possible reaction pathways are excluded from the analysis.
The experimental signal at $\text{KER} \approx 36$ eV and $\Theta\approx 153^\circ$ corresponds to static $\text{CH}_2\text{I}_2$ (zero pump-probe delay) and shows reasonable agreement with the theoretical prediction. As the pump-probe delay increases, the KER decreases due to the increasing C–I and I–I distances. Meanwhile, the angle $\Theta$ first increases slightly and then decreases by about $5-10^\circ$, which is attributed to the onset of rotation of the $\text{CH}_2\text{I}$ radical during $\text{CH}_2\text{I}+\text{I}$ dissociation.
Simulations using the $2\Gamma_1$ state show better agreement with the experiment than those using the $6\Gamma_2$ state. The signals corresponding to the $2\Gamma_1$ state lie within the statistical error range of the experimental data, while those corresponding to the $6\Gamma_2$ state exhibit lower KER values and larger $\Theta$. These results may suggest that the molecule is more likely to transition to the $2\Gamma_1$ state upon UV excitation.

\section{Conclusion}
In summary, we have developed a theoretical model to simulate the UV-induced photodissociation of diiodomethane and the subsequent three-fragment Coulomb explosion, in analogy to a recent pump-probe experiment using coincident ion momentum imaging to probe real-time molecular structures\cite{anbu2025}.
We select only two internal DOF as the reaction coordinates to describe the C–I bond breaking and the $\text{CH}_2\text{I}$ rotation during photodissociation, and only three internal DOF to describe the coincident $\text{CH}_2^{+} + \text{I}^{2+} + \text{I}^{2+}$ three-ion fragmentation in the subsequent Coulomb explosion. Based on high-level \textit{ab initio} electronic structure calculations, two-dimensional PESs of the valence excited states of the neutral molecule are built using spline interpolation, and the three-dimensional ground-state PES of the fivefold-charged cation is constructed using the many-body expansion approach.
The photodissociation pathways and the final momenta of the ionic fragments are obtained by solving the EOM on these reduced-dimensional PESs. The photodissociation  pathways for the $2\Gamma_1$ and $6\Gamma_2$ states, which lead to the $\text{CH}_2\text{I}+\text{I}(^2\text{P}_{3/2})$ and $\text{CH}_2\text{I}+\text{I}^*(^2\text{P}_{1/2})$ channels, respectively, show overall agreement with previous AIMD simulations in terms of the time-dependent C–I and I–I pair distances and the I–C–I bond angle. These results also reveal that the $\text{CH}_2\text{I}$ radical rotates with a period of approximately 340 fs during photodissociation.
The theoretical KER obtained using the full ionic potential shows better agreement with experimental signals than that obtained using a pure Coulomb potential, reflecting the importance of non-Coulombic effects during ion fragmentation. The correlation between KER and $\Theta$ (the angle between the momenta of the two $\text{I}^{2+}$ ions) is also compared with experimental signals, which confirms the unpumped static $\text{CH}_2\text{I}_2$ state and the $\text{CH}_2\text{I} + \text{I}$ dissociation channels. 


\begin{acknowledgments}
The author thanks Daniel Rolles, Artem Rudenko, B. D. Esry, and Yusong Liu for helpful discussions at the early stage of this study. The author thanks Anbu Selvam Venkatachalam for providing the experimental data shown in Fig. \ref{fig5} and \ref{fig6}. The author thanks Bin Zhao and all other group members for their hospitality during the author's visiting at Southern University of Science and Technology.

\end{acknowledgments}

\section*{Data Availability Statement}
All data supporting the findings of this study are available from the corresponding author upon reasonable request.

\nocite{*}
\bibliography{aipsamp}

\end{document}